\documentclass[pra,twocolumn,showpacs,superscriptaddress,floatfix]{revtex4}

\usepackage{amsmath}
\usepackage{times}
\usepackage{txfonts}
\usepackage{graphicx}

\mathchardef\ogon="012C%
\newcommand{\as}{a\kern-0.22em\lower.40ex\hbox{$_{\ogon}$}}

\begin{document}

\title{Heating and atom loss during upward ramps of Feshbach resonance levels 
in Bose-Einstein condensates}
\author{Thorsten K\"{o}hler} 
\affiliation{Clarendon Laboratory, Department of Physics, 
University of Oxford, Parks Road, Oxford, OX1 3PU, United Kingdom}
\author{Krzysztof G{\'o}ral}
\affiliation{Clarendon Laboratory, Department of Physics, 
University of Oxford, Parks Road, Oxford, OX1 3PU, United Kingdom}
\affiliation{Center for Theoretical Physics, Polish Academy of 
Sciences, Aleja Lotnik\'ow 32/46, 02-668 Warsaw, Poland}
\author{Thomas Gasenzer} 
\affiliation{Institut f\"ur Theoretische Physik, Universit\"at Heidelberg, 
Philosophenweg 16, 69120 Heidelberg, Germany}

\begin{abstract}
The production of pairs of fast atoms leads to a pronounced loss of atoms
during upward ramps of Feshbach resonance levels in dilute Bose-Einstein 
condensates. We provide comparative studies on the formation of these bursts 
of atoms containing the physical predictions of several theoretical 
approaches at different levels of approximation. We show that despite 
their very different description of the microscopic binary physics during the 
passage of a Feshbach resonance, all approaches lead to virtually the same 
prediction on the total loss of condensate atoms, provided that the ramp of 
the magnetic field strength is purely linear. We give the reasons for this 
remarkable insensitivity of the remnant condensate fraction to the 
microscopic physical processes and compare the theoretical predictions 
with recent Feshbach resonance crossing experiments on $^{23}$Na and $^{85}$Rb.
 
\end{abstract}
\date{\today}
\pacs{03.75.Kk, 34.50.-s, 05.30.-d}
\maketitle

\section{Introduction}
Since their first experimental demonstration 
\cite{Inouye98}, 
magnetic field tunable interactions have proven their
potential and usefulness in the physics of quantum degenerate gases. This 
technique takes advantage of the Zeeman effect in the 
electronic energy levels of the atoms to manipulate their binary collision 
properties in the vicinity of a Feshbach resonance. For several atomic species 
it is now possible to change the scattering length by orders of
magnitude, and even to reverse its sign, making the atoms either repel
or attract one another. One of the impressive applications of this technique 
was the successful creation of stable 
$^{85}$Rb Bose-Einstein condensates with thousands of atoms 
\cite{Cornish00}, that would otherwise collapse if consisting of more 
than about 80 atoms. These experiments have now been refined to 
associate pairs of atoms to ultracold molecules and probe their coherence 
properties 
\cite{Donley02,Claussen03,Regal03,Herbig03,Duerr03,Strecker03,Cubizolles03,%
 JochimPRL03,Regal04,Xu03,Greiner03,JochimScience03,Mukaiyama03,Zwierlein03,%
 Barenstein04,Regal04-2,Zwierlein04,Barenstein04-2}. The dissociation
energies of the highly excited dimer molecules have been measured
accurately down to several kHz \cite{Donley02,Claussen03,Regal03}. 

Early attempts to associate Feshbach molecules in Bose-Einstein 
condensates \cite{Stenger99} relied upon linear ramps of the magnetic field 
strength across the resonance position, in which the resonance level 
(see, e.g., \cite{Mies00,GKGTJ03}) was swept upward in energy \cite{footnote:upward}. Subsequent 
theoretical studies \cite{Abeelen99,Mies00} have shown that these particular
ramps lead to a heating of the cloud rather than to molecular production.
The condensate atoms were lost into correlated pairs of atoms with a 
comparatively high velocity of their relative motion. These atom pairs were
also produced in the interferometric experiments of Ref.~\cite{Donley02},
involving a special sequence of magnetic field pulses, and directly detected
as a dilute burst of atoms with a spread far beyond the size of the remnant
Bose-Einstein condensate. The theoretical studies in 
Refs.~\cite{Kokkelmans02,Mackie02,KGB02} have confirmed this interpretation and
provided a quantitative understanding of the number of burst atoms observed in 
Ref.~\cite{Donley02} as well as their measured energy spectrum \cite{KGB02}. 

Several approaches have been proposed to describe the conversion of condensate 
atoms into bound or continuum molecular states
(see, e.g., \cite{Drummond98,Timmermans99,Abeelen99,Yurovsky99,Mies00,KGB02,Kokkelmans02,Mackie02,Koehler02,Duine03,Naidon03,Yurovsky03,GKGTJ03}). 
Early approaches are based either 
on mean field theory, with an exchange of particles between the atomic 
condensate and the Feshbach resonance level 
(e.g.~\cite{Drummond98,Timmermans99,Abeelen99}), 
or on a two-body configuration interaction picture, extended in such a way 
that it includes the Bose stimulation of the transition between free atom pairs
and the resonance level \cite{Mies00}. While these approaches effectively 
include only two energy levels of the atom pairs, i.e.~the zero energy 
scattering  state and the Feshbach resonance level \cite{Mies00,GKGTJ03}, the 
interferometric experiments in Ref.~\cite{Donley02} have clearly revealed 
that a complete description of the molecular association requires many-body
approaches, which treat the two-body collision dynamics including all 
states in a non-perturbative manner \cite{Koehler02,Kokkelmans02,KGB02}. 

In this paper we use the beyond mean field approach of 
Refs.~\cite{Koehler02,KGB02} to describe the atom loss in upward ramps of a 
Feshbach resonance level in a Bose-Einstein condensate and compare the results 
with the simpler two level mean field approach of 
Refs.~\cite{Drummond98,Timmermans99,Abeelen99}. Our studies involve the 
Feshbach resonance crossing experiments of Refs.~\cite{Stenger99} and 
\cite{Cornish00}. In the experiments \cite{Stenger99} a Bose-Einstein 
condensate of $^{23}$Na in the $(F=1,m_F=+1)$ hyperfine state was exposed to a 
magnetic field swept across the Fesh\-bach resonances at 853 and 907 G
\cite{Inouye98}, respectively. In Ref.~\cite{Cornish00} the magnetic field 
strength was swept across the 155 G resonance of a gas of $^{85}$Rb atoms, 
which was condensed in the $(F=2,m_F=-2)$ hyperfine state. In all cases 
dramatic losses of atoms were observed, which increased with decreasing ramp 
speed. 
 
The widths of the $^{23}$Na Feshbach resonances have been predicted to be 
1 G (at 907 G) and 0.01 G (at 853 G), respectively 
\cite{Inouye98}. The $^{85}$Rb 155 G Feshbach resonance has been 
characterised very accurately in several experimental and theoretical studies 
\cite{Donley02,Roberts01,vanKempen02,Claussen03}, and its width is predicted
to be 11 G. We shall show on the basis of binary collision physics that 
the microscopic dynamics caused by the ramps of the magnetic field strength 
is decisively different between the $^{23}$Na experiments \cite{Stenger99}  
and the experiments \cite{Cornish00} with the broad $^{85}$Rb Feshbach 
resonance, although the direction of these resonance crossings excludes the 
adiabatic association of atoms to molecules (see, e.g., \cite{Mies00,GKGTJ03}) 
in both cases. We shall show that a two level picture is particularly
inappropriate to describe the Feshbach resonance crossing in $^{85}$Rb 
\cite{Cornish00}. Our studies reveal, however, that both the first order
microscopic quantum dynamics approach of Refs.~\cite{Koehler02,KGB02} and the 
two level mean field theory of Refs.~\cite{Drummond98,Timmermans99,Abeelen99} 
predict largely the same atom loss for the experiments in 
Refs.~\cite{Stenger99} and \cite{Cornish00}. We show that this remarkable 
insensitivity of the remnant condensate fraction with respect to the 
description of the microscopic binary collision physics is characteristic for 
{\em linear} ramps of the magnetic field strength and provide the essential 
physical parameters that determine the losses. We show furthermore that our 
findings are quite general as they also apply to the dissociation spectra of 
Feshbach molecules in upward ramps of the Feshbach resonance level, which are 
widely used for the purpose of molecular detection 
(see, e.g., \cite{Herbig03,Duerr03,Mukaiyama03,Duerr04}). We note,
however, that the interferometric experiments of
Refs.~\cite{Donley02,Claussen03} provide an example of
a sequence of  magnetic field pulses, which as a whole varies
non-linearly in time. These experiments indeed reveal details of  the
microscopic binary collision physics beyond
universal considerations \cite{Kokkelmans02,KGB02,KGTKKB04}.

\section{Microscopic quantum dynamics approach}
In general, the interatomic interaction depends on the relative coordinates of
the atoms and couples all their hyperfine energy levels, labelled in 
the following by Greek indices. A trapped gas of Bose atoms is then described 
by the many-body Hamiltonian: 
\begin{widetext}
\begin{align}
  H=\sum_\mu\int d\mathbf{x} \ \psi_\mu^\dagger(\mathbf{x})
  H^\mathrm{1B}_\mu(B)\psi_\mu(\mathbf{x})
  +\frac{1}{2}\sum_{\mu\nu\kappa\lambda}
  \int d\mathbf{x} \int d\mathbf{y} \ \psi^\dagger_\mu(\mathbf{x})
  \psi^\dagger_\nu(\mathbf{y})V_{\{\mu\nu\},\{\kappa\lambda\}}
  (\mathbf{x}-\mathbf{y})
  \psi_\kappa(\mathbf{x})\psi_\lambda(\mathbf{y}).
  \label{Hamiltonian}
\end{align}
\end{widetext}
Here $H^\mathrm{1B}_\mu(B)=-\hbar^2\nabla^2/2m+V_\mu^\mathrm{trap}+
E_\mu^\mathrm{hf}(B)$
is the Hamiltonian of a single atom in the hyperfine state $\mu$ containing 
the kinetic energy, the trap potential and the hyperfine energy
that depends on the magnetic field strength $B$ due to the Zeeman effect,
and $m$ is the atomic mass. $V_{\{\mu\nu\},\{\kappa\lambda\}}(\mathbf{r})$ 
is the potential associated with the asymptotic incoming and outgoing binary 
scattering channels $\{\kappa\lambda\}$ and $\{\mu\nu\}$, which depends on 
the relative coordinates $\mathbf{r}=\mathbf{x}-\mathbf{y}$. The field 
operators obey the Bose commutation relations 
$[\psi_\mu(\mathbf{x}),\psi_\nu^\dagger(\mathbf{y})]=\delta_{\mu\nu}
\delta(\mathbf{x}-\mathbf{y})$ and 
$[\psi_\mu(\mathbf{x}),\psi_\nu(\mathbf{y})]=0$.

All physical quantities of a gas are determined by correlation 
functions, i.e., expectation values (denoted by $\langle\cdots\rangle$)
of normal ordered products of field operators in the state at time $t$. 
Quantities of particular interest are the mean field
$\Psi_\mu(\mathbf{x},t)=\langle \psi_\mu(\mathbf{x}) \rangle_t$, 
the pair function $\Phi_{\mu\nu}(\mathbf{x},\mathbf{y},t)=\langle
\psi_\nu(\mathbf{y}) \psi_\mu(\mathbf{x}) \rangle_t - 
\Psi_\nu(\mathbf{y},t) \Psi_\mu(\mathbf{x},t)$ and the one-body density
matrix of the non-condensed fraction 
$\Gamma_{\mu\nu}(\mathbf{x},\mathbf{y},t)=\langle
\psi_\nu^{\dagger}(\mathbf{y}) \psi_\mu(\mathbf{x}) \rangle_t - 
\Psi_\nu^*(\mathbf{y},t)\Psi_\mu(\mathbf{x},t)$. We have shown in 
Refs.~\cite{Koehler02,KGB02} how the exact infinite hierarchy of dynamic 
equations for correlation functions can be truncated to any desired 
degree of accuracy. In the first order approach
\cite{KGB02}
the mean field and the pair function are determined by the coupled 
dynamic equations \cite{GKGTJ03}:
\begin{widetext}
\begin{align}
  \label{meanfield}
  i\hbar\frac{\partial}{\partial t}\Psi_\mu(\mathbf{x},t)=&H_\mu^\mathrm{1B}(B)
  \Psi_\mu(\mathbf{x},t)+
  \sum_{\nu\kappa\lambda}\int d\mathbf{y} \ \Psi_\nu^*(\mathbf{y},t)
  V_{\{\mu\nu\},\{\kappa\lambda\}}(\mathbf{x}-\mathbf{y})
  \left[
    \Phi_{\kappa\lambda}(\mathbf{x},\mathbf{y},t)+
    \Psi_\kappa(\mathbf{x},t)\Psi_\lambda(\mathbf{y},t)
    \right],\\
  i\hbar\frac{\partial}{\partial t}\Phi_{\mu\nu}(\mathbf{x},\mathbf{y},t)
  =&\sum_{\kappa\lambda}
  \left[
    H^\mathrm{2B}_{\{\mu\nu\},\{\kappa\lambda\}}(B)
    \Phi_{\kappa\lambda}(\mathbf{x},\mathbf{y},t)+
    V_{\{\mu\nu\},\{\kappa\lambda\}}(\mathbf{x}-\mathbf{y})
    \Psi_\kappa(\mathbf{x},t)\Psi_\lambda(\mathbf{y},t)
    \right].
  \label{pairfunction}
\end{align}
\end{widetext}
Here $H_{\{\mu\nu\},\{\kappa\lambda\}}^\mathrm{2B}(B)=
[H_\mu^\mathrm{1B}(B)+H_\nu^\mathrm{1B}(B)]\delta_{\kappa\mu}
\delta_{\lambda\nu}+V_{\{\mu\nu\},\{\kappa\lambda\}}$ is the Hamiltonian 
matrix of two atoms. All the other correlation functions can be expressed in 
terms of the mean field and the pair function \cite{KGB02}. The coupled
equations (\ref{meanfield}) and (\ref{pairfunction}) are the most general 
form of the first order microscopic quantum dynamics approach of 
Refs.~\cite{Koehler02,KGB02}, but they also include the two level mean field 
approach of Refs.~\cite{Drummond98,Timmermans99,Abeelen99} in the Markov
approximation \cite{GKGTJ03}. Higher order approximations \cite{TK02} in the 
general truncation scheme \cite{Koehler02} include few-body inelastic loss 
phenomena, like three-body recombination \cite{footnote:inelastic}. We shall 
focus in this work on the dominant two-body scattering in the presence of the 
surrounding Bose gas.

\section{Resonance enhanced binary collisions}
\subsection{General properties of Feshbach resonances in ultracold collisions}
\begin{figure}[htb]
\begin{center}
\includegraphics[width=\columnwidth,clip]{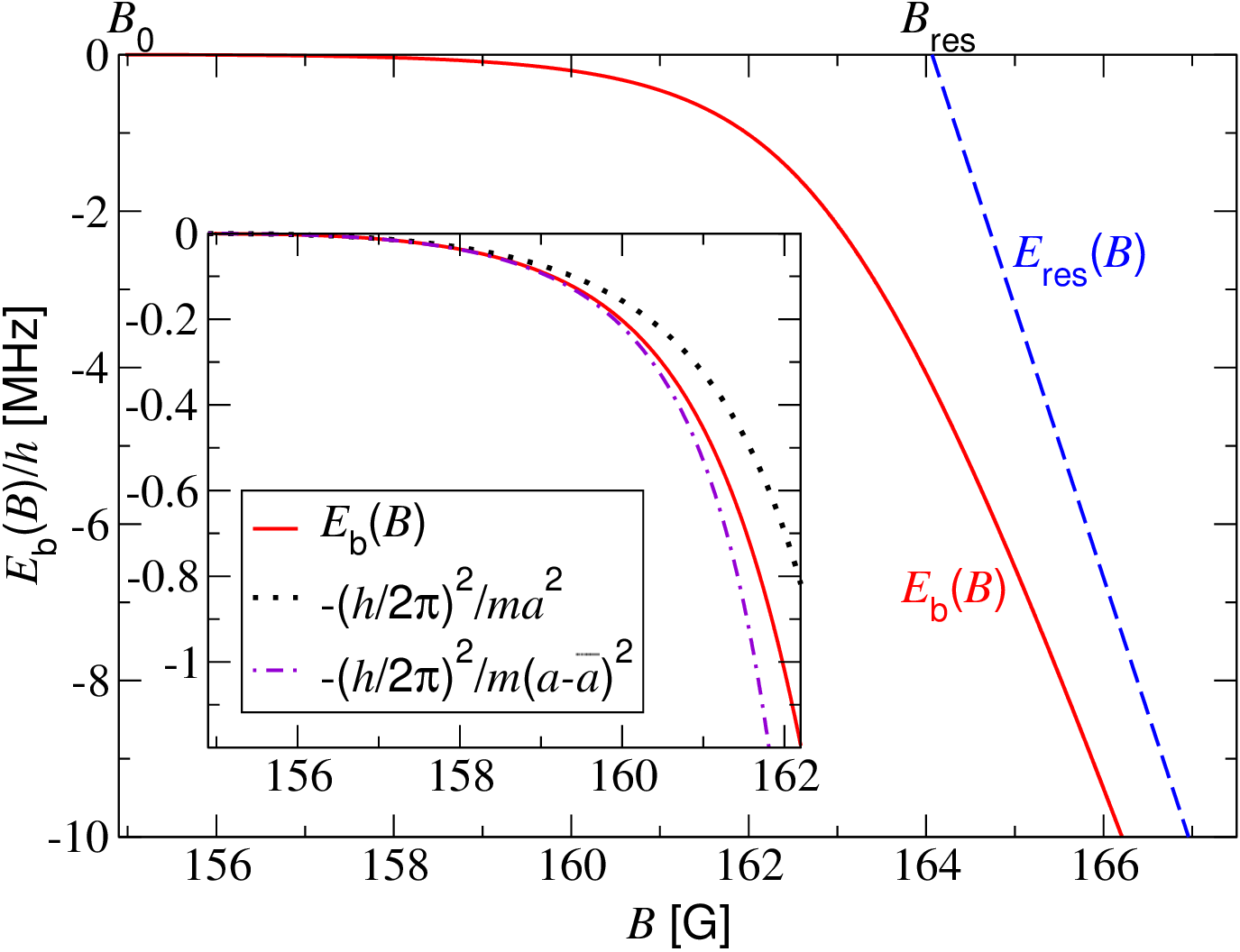}
\caption{Magnetic field dependence of the binding energy $E_\mathrm{b}(B)$
of the highest excited multichannel vibrational bound state $\phi_\mathrm{b}$ 
of $^{85}$Rb$_2$ (solid curve) and the energy $E_\mathrm{res}(B)$ of the 
closed channel resonance state $\phi_\mathrm{res}(r)$ (dashed curve). 
$E_\mathrm{res}(B)$ depends linearly on $B$ with 
$h^{-1}dE_\mathrm{res}/dB=-3.46 \ \mathrm{MHz}/\mathrm{G}$.
The inset shows $E_\mathrm{b}(B)$ (solid curve) as compared to the 
universal asymptotic formula $-\hbar^2/ma^2$ for the near resonant
binding energy (dotted curve) and its first improvement 
\protect\cite{Gribakin93} that accounts for the length scale set by the van 
der Waals interaction in a single channel picture (dotted dashed curve).}
\label{fig:E_b85Rb}
\end{center}
\end{figure}

We denote the open $s$ wave binary scattering channel of two asymptotically 
free $^{23}$Na condensate atoms in the $(F=1,m_F=1)$ state as the $\{1,1\}$ 
open channel (or simply as the open channel) with an associated background  
scattering potential $V_{\{1,1\},\{1,1\}}(\mathbf{r})=V_\mathrm{bg}(r)$,
where $r$ is the internuclear distance. Analogously, for $^{85}$Rb the 
background scattering potential is denoted by
$V_{\{-2,-2\},\{-2,-2\}}(\mathbf{r})=V_\mathrm{bg}(r)$. Throughout this
paper we shall choose the zero of energy as the dissociation threshold of 
$V_\mathrm{bg}(r)$, i.e.~$V_\mathrm{bg}(r)\underset{r\to\infty}{\sim}0$. 
Figure \ref{fig:E_b85Rb} shows, for the case of the 
$^{85}$Rb Feshbach resonance, that in the vicinity of the range of magnetic 
field strengths (from 162.2 to 132 G) in Ref.~\cite{Cornish00} the energy 
$E_\mathrm{res}(B)$ of a closed channel vibrational state 
(a Feshbach resonance level) $\phi_\mathrm{res}(r)$ crosses zero (at 
$B_\mathrm{res}=164.1$ G). The resulting strong coupling between the 
open and the closed channel is accompanied by the emergence of a highly excited
multichannel vibrational molecular bound state $\phi_\mathrm{b}$, whose 
binding energy $E_\mathrm{b}(B)$ vanishes at the measurable position of the 
Feshbach resonance of $B_0=154.9$ G. When the magnetic field strength is 
swept across the resonance from above, i.e.~the energy of the Feshbach 
resonance level $\phi_\mathrm{res}(r)$ is tuned upward, $\phi_\mathrm{b}$ 
thus ceases to exist (see Fig.~\ref{fig:E_b85Rb}). The $s$ wave scattering 
length of two asymptotically free condensate atoms, therefore, has a 
singularity at $B_0$. The emergence of a multichannel diatomic bound state, 
which causes the singularity of the scattering length at $B_0$, is the 
decisive feature of all Feshbach resonances in ultracold collision physics.

\subsection{Two channel approach}
The coupling between the $\{-2,-2\}$ open channel of $^{85}$Rb and other open 
channels, leading, e.g., to inelastic two-body losses, is weak 
\cite{footnote:inelastic}. In the case of $^{23}$Na the condensate atoms are 
in the electronic ground state and inelastic two-body losses are ruled out 
from the outset. It thus turns out that for the Feshbach resonances of both 
species the Hamiltonian of the relative motion of two atoms in free space can 
be reduced to just two channels \cite{Mies00,TKTGPSJKB03}:
\begin{align}
  \label{H_2B}
  H_\mathrm{2B}(B)
  &=\ -\frac{\hbar^2}{m}\nabla^2
  \left[|\mathrm{bg}\rangle\langle\mathrm{bg}|
    +|\mathrm{cl}\rangle\langle\mathrm{cl}|\right]
  +V_\mathrm{2B}(B),\\
  V_\mathrm{2B}(B)
  &=\ (|\mathrm{bg}\rangle,|\mathrm{cl}\rangle)
      \left(\begin{array}{cc}
      V_\mathrm{bg}(r) & 
      W(r) \\
      W(r)  & 
      V_\mathrm{cl}(B,r)
      \end{array}\right)
      \left(\!\!\begin{array}{c}
      \langle\mathrm{bg}|\\ 
      \langle\mathrm{cl}|
      \end{array}\!\!\right).
  \label{V_2B}
\end{align}
Here $|\mathrm{bg}\rangle$ is the product of hyperfine states associated with 
the open channel, $|\mathrm{cl}\rangle$ is the superposition of products of 
atomic hyperfine states associated with the strongly coupled closed channel 
\cite{Mies00}, and $W(r)$ couples the two channels. $V_\mathrm{cl}(B,r)$ is 
the closed channel potential that supports the resonance state 
$\phi_\mathrm{res}(r)$, i.e.
\begin{equation} 
  \left[-\hbar^2\nabla^2/m+V_\mathrm{cl}(B,r)\right]\phi_\mathrm{res}(r)=
  E_\mathrm{res}(B)\phi_\mathrm{res}(r). 
\end{equation}
As only the resonance state couples   
strongly to the open channel, $-\hbar^2\nabla^2/m+V_\mathrm{cl}(B,r)$ can be 
replaced by the one dimensional Hamiltonian 
$|\phi_\mathrm{res}\rangle E_\mathrm{res}(B) \langle\phi_\mathrm{res}|$
to an excellent approximation \cite{GKGTJ03}. Within the limited range of 
magnetic field strengths under consideration the energy associated with the 
resonance state is linear in the magnetic field strength, i.e.
\begin{equation}
  E_\mathrm{res}(B)=\left[\frac{dE_\mathrm{res}}{dB}(B_\mathrm{res})\right]
  (B-B_\mathrm{res})\, .
\end{equation}
Here $dE_\mathrm{res}/dB$ denotes the difference between the magnetic
moment of the resonance state and an atom pair in the open channel, and is
usually referred to as the slope of the resonance.

\subsection{Effective low energy potentials}
\begin{figure}[htb]
\begin{center}
\includegraphics[width=\columnwidth,clip]{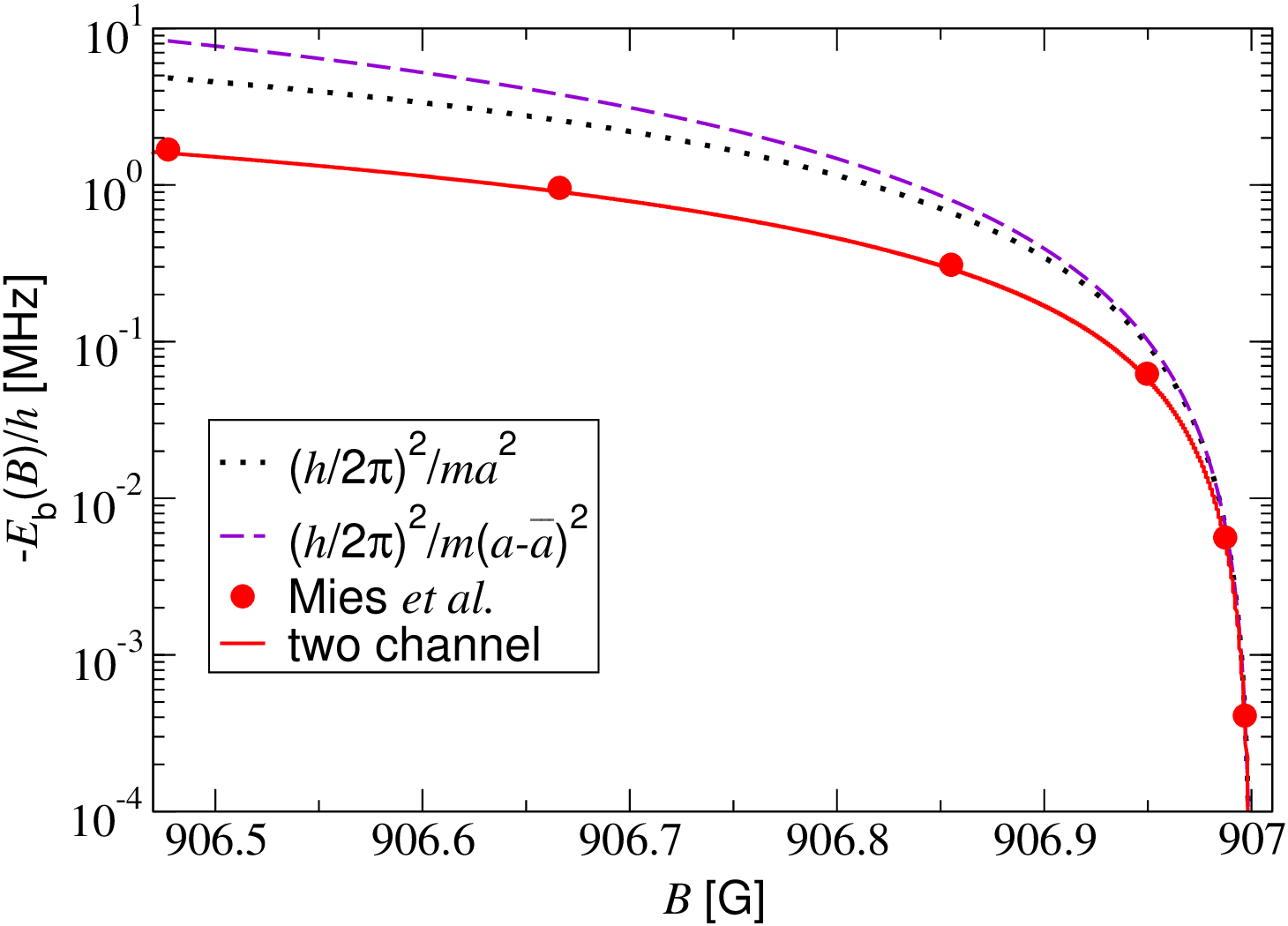}
\caption{Magnetic field dependence of the modulus of the binding energy of
the highest excited multichannel vibrational bound state of $^{23}$Na$_2$ in 
the vicinity of the 907 G Feshbach resonance. The figure compares the 
binding energies supported by the two channel low energy potential matrix of 
Ref.~\cite{GKGTJ03} and this work (solid curve) with the highly accurate two 
channel model of Ref.~\cite{Mies00} (filled circles). The dotted curve 
indicates the universal asymptotic formula $E_\mathrm{b}=-\hbar^2/ma^2$ and
the dashed curve is the Gribakin and Flambaum \protect\cite{Gribakin93}
estimate that accounts for the length scale set by the van der Waals 
interaction in a single channel picture. We note that the binding energies are 
given on a logarithmic scale.}
\label{fig:E_bNa907G}
\end{center}
\end{figure}

The typical de Broglie wave lengths of the condensate atoms are much larger
than the length scales set by the spatial extent of the binary interactions. 
The details of the potentials are, therefore, not resolved in the binary 
collisions. Consequently, we can use the approach of Ref.~\cite{GKGTJ03} whose
parameters for the $^{23}$Na and $^{85}$Rb Feshbach resonances we shall briefly
summarise. For the potential energy operator $V_\mathrm{bg}$ we use the 
separable form
\begin{equation}
  V_\mathrm{bg}=|\chi_\mathrm{bg}\rangle\xi_\mathrm{bg}
  \langle\chi_\mathrm{bg}|, 
\end{equation}
with the convenient Gaussian {\em ansatz} 
\begin{equation}
  \chi_\mathrm{bg}(r)=\frac{e^{-r^2/2\sigma_\mathrm{bg}^2}}{(\sqrt{2\pi}
  \sigma_\mathrm{bg})^3}, 
\end{equation}
as introduced in Ref.~\cite{KGB02} to describe the low energy background 
scattering of $^{23}$Na \cite{Mies00} and $^{85}$Rb 
\cite{Donley02,Roberts01,vanKempen02,Claussen03}. 
The background scattering potential $V_\mathrm{bg}$ then recovers the 
background scattering length of the atoms in the vicinity of the respective 
resonance, as well as the length scale 
\begin{equation}
  l_\mathrm{vdW}=\frac{1}{2}(mC_6/\hbar^2)^{1/4}
  \label{lvdW}
\end{equation}
set by the van der Waals interaction $-C_6/r^6$, which is determined by the 
dispersion coefficient $C_6$. Table \ref{tab:resparams} provides the relevant 
physical parameters associated with the Feshbach resonances considered in this 
paper, and Table \ref{tab:seppotparams} shows the parameters $\xi_\mathrm{bg}$ 
and $\sigma_\mathrm{bg}$ characterising the low energy background 
scattering potential. To recover the widths $\Delta B$ of 
the resonances and the shifts $B_0-B_\mathrm{res}$ (see Fig.~\ref{fig:E_b85Rb})
as given in Table \ref{tab:resparams} it is sufficient to specify 
$W(r)\phi_\mathrm{res}(r)$. We have used the {\em ansatz}
\begin{equation} 
W(r)\phi_\mathrm{res}(r)=\zeta\frac{e^{-r^2/2\sigma^2}}{(\sqrt{2\pi}\sigma)^3}.
\end{equation}
The parameters $\zeta$ and $\sigma$ can be obtained from 
Table \ref{tab:seppotparams}. The scattering length is then exactly given by 
the formula 
\begin{equation}
  \label{scattlength}
  a(B)=a_\mathrm{bg}\left(1-\frac{\Delta B}{B-B_0}\right). 
\end{equation}
The binding energies $E_\mathrm{b}(B)$ in Fig.~\ref{fig:E_b85Rb} have been 
obtained with this low energy potential matrix $V_\mathrm{2B}(B)$ by
\begin{equation} 
  H_\mathrm{2B}\phi_\mathrm{b}=E_\mathrm{b}\phi_\mathrm{b}. 
\end{equation}
Figure \ref{fig:E_bNa907G} shows a comparison of $E_\mathrm{b}(B)$, for the
907 G Feshbach resonance of $^{23}$Na, between the low energy 
potentials in this work and the highly accurate two channel model of 
Ref.~\cite{Mies00}. The associated two-body time evolution operator 
$U_\mathrm{2B}(t,\tau)$ can be obtained by relatively simple means 
\cite{KGB02}, even for the time dependent magnetic field strengths in 
Refs.~\cite{Stenger99,Cornish00}. $U_\mathrm{2B}(t,\tau)$ will serve as an 
input to the solution of the coupled equations (\ref{meanfield}) and 
(\ref{pairfunction}). 

\begin{center}
\begin{table*}[tb]
\begin{tabular}
       {l@{\quad}rl
         @{\quad}rl
         @{\quad}rl
         @{\quad}rl
         @{\quad}rl
         @{\quad}rl
         @{\quad}rl}
Element & $a_\mathrm{bg}$       &[$a_\mathrm{B}$]
        & $C_6$                 &[a.u.]
        & $E_{-1}$              &[MHz $h$]
        & $B_0$                 &[G]
         & $\Delta B$           &[G]
        & $B_0-B_\mathrm{res}$  &[G]
        & $\partial E_\mathrm{res}/\partial B$  &[MHz h/G]\\[0.5ex]
\hline
$^{23}$Na & 63.9        &\cite{Mies00}
           & 1561       &\cite{Kharchenko97}
          & 207.8       &\cite{Julienneprivcomm}
          & 853         &\cite{Stenger99}
          & 0.01        &\cite{Mies00}
           & -5.85 $\times$10$^{-3}$     &\cite{JTK03}
           & 5.24       &\cite{Mies00}\\
           & 62.8       &\cite{Mies00}
           & 1561       &\cite{Kharchenko97}
          & 207.8       &\cite{Julienneprivcomm}
          & 907         &\cite{Stenger99}
          & 1.0         &\cite{Mies00}
           & -0.55      &\cite{JTK03}
           & 5.24       &\cite{Mies00}\\
$^{85}$Rb & -450        &\cite{Donley02}
           & 4660       &\cite{Roberts01}
           &            &
          & 155         &\cite{Roberts01}
           & 11.0       &\cite{Roberts01}
           & -9.2       &\cite{JTK03}
           & -3.46      &\cite{Mies00}
\end{tabular}
\caption{\label{tab:resparams}Parameters characterising the Feshbach 
resonances ($a_\mathrm{Bohr}=0.052918$ nm; $C_6$: 1 
a.u.$=0.095734\times 10^{-24}\,$J nm$^6$).}
\end{table*}

\begin{table*}[tb]
\begin{tabular}{l@{\qquad}c@{\qquad}c@{\qquad}c@{\qquad}c@{\qquad}c}
Element & $B_0$ [G] 
        & $\xi_\mathrm{bg}$ [$4\pi\hbar^2a_\mathrm{Bohr}/m$] 
	& $\sigma_\mathrm{bg}$ [$a_\mathrm{Bohr}$] 
	& $\zeta$ [GHz $h\,a_\mathrm{Bohr}^{3/2}$] 
	& $\sigma$ [$a_\mathrm{Bohr}$] \\[0.5ex]
\hline
$^{23}$Na & 853 & -197 & 27.2 & -3.96 & 14.5 \\ 
          & 907 & -186 & 26.5 & -44.0 & 17.4 \\ 
$^{85}$Rb & 155 & -99.3& 71.9 &  19.0 & 52.4
\end{tabular}
\caption{\label{tab:seppotparams}Parameters determining the low energy 
two channel potential matrix.}
\end{table*}
\end{center}

\begin{figure}[htb]
\begin{center}
\includegraphics[width=\columnwidth,clip]{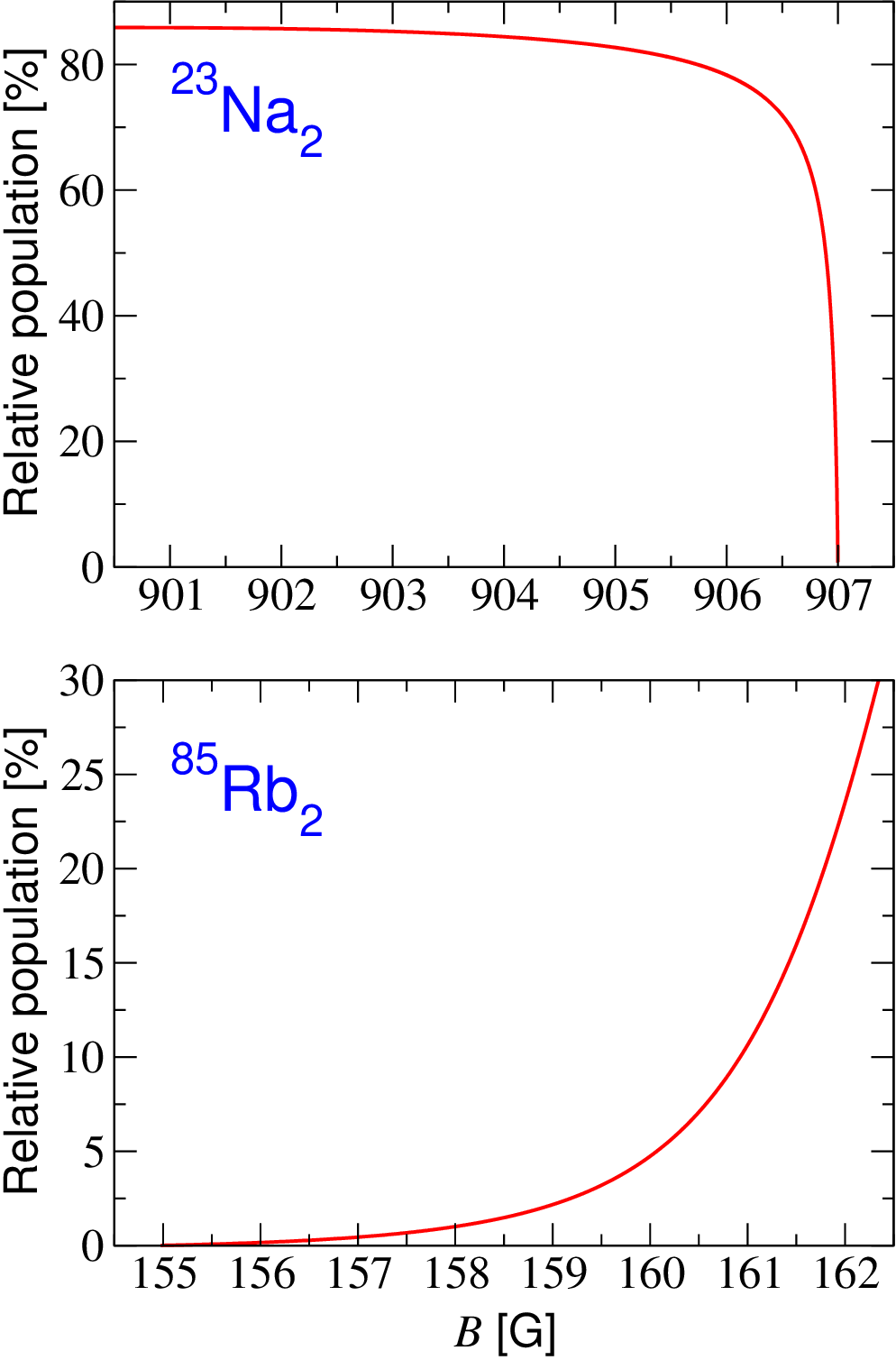}
\caption{Magnetic field dependence of the population of the closed channel 
  component of the highest excited vibrational bound state $\phi_\mathrm{b}$  
  associated with the $^{23}$Na 907 G and the $^{85}$Rb 155 G 
  Feshbach resonance. The populations were determined from the two channel 
  Hamiltonian (\ref{H_2B}).}
\label{fig:populationcl}
\end{center}
\end{figure}

\subsection{Differences between the $^{23}$Na and $^{85}$Rb Feshbach 
resonances}
\label{subsec:differencesbetweenresonances}
Although the $^{23}$Na and $^{85}$Rb Feshbach resonances considered in this 
paper can all be described by a low energy two channel Hamiltonian of the form 
of Eq.~(\ref{H_2B}), with a single resonance state, the multichannel bound 
states which cause the singularities of the scattering length differ decisively
among the resonances within comparable ranges of the magnetic field strength. 
The differences are illustrated in Fig.~\ref{fig:populationcl}, which shows the
admixture $|\langle\phi_\mathrm{res},\mathrm{cl}|\phi_\mathrm{b}\rangle|^2$ of 
the closed channel state to the two channel vibrational molecular bound state 
$\phi_\mathrm{b}$ of the Hamiltonian (\ref{H_2B}). For $^{85}$Rb the 
admixture stays small ($<30$ \%) over the entire range of experimental 
\cite{Cornish00} magnetic field strengths from 154.9 to 162.2 G on the 
high field side of the Feshbach resonance (cf., also, Ref.~\cite{TKTGPSJKB03}).
Even for the broader ($\Delta B=1$ G) resonance of $^{23}$Na at 907 G, 
however, the multichannel bound state $\phi_\mathrm{b}$ is transferred 
immediately into the closed channel resonance state $\phi_\mathrm{res}(r)$, 
with an admixture of more than 80 \% within the resonance width. The single 
channel nature of the $^{85}$Rb 155 G Feshbach resonance is also visible in
the binding energies in Fig.~\ref{fig:E_b85Rb} because the near resonant single
channel Gribakin and Flambaum estimate
\cite{Gribakin93,KGB02}
\begin{equation}
  E_\mathrm{b}=-\hbar^2/m(a-\bar{a})^2
  \label{E_bGF}
\end{equation} 
is applicable, where
\begin{equation}
  \bar{a}=\frac{1}{\sqrt{2}}\frac{\Gamma(3/4)}{\Gamma(5/4)}l_\mathrm{vdW} 
\end{equation}
is the mean scattering length \cite{Gribakin93}, which accounts for the long 
range behaviour $-C_6/r^6$ of the background scattering potential in terms of 
the van der Waals length in Eq.~(\ref{lvdW}). For the 907 G Feshbach 
resonance of $^{23}$Na, however, Fig.~\ref{fig:E_bNa907G} shows that 
Eq.~(\ref{E_bGF}) is even less accurate than the universal near resonant 
estimate \cite{Donley02,TKTGPSJKB03,GKGTJ03} $E_\mathrm{b}=-\hbar^2/ma^2$
of the binding energy.

The reason for the single channel nature of the $^{85}$Rb 155 G Feshbach 
resonance is the large shift of $B_0-B_\mathrm{res}=-9.2$ G 
(see Fig.~\ref{fig:E_b85Rb}), which leads to a large detuning of 
$E_\mathrm{res}(B_0)/h=32$ MHz at the resonance position. The naive physical
two level picture, in which the resonance level $\phi_\mathrm{res}(r)$ is swept
across the zero energy level of the background scattering potential at the 
measurable position $B_0$ of the Feshbach resonance, and is then transferred 
to a bound molecular state $\phi_\mathrm{b}$ at negative energies 
$E_\mathrm{res}<0$ (see, e.g., Ref.~\cite{JTK03}), is therefore  
inapplicable. In fact, the resonance state does not even cross the threshold 
within the entire range of experimental magnetic field strengths 
(from 162.2 to 132 G) in Ref.~\cite{Cornish00}. 

\section{Many-body approaches to Feshbach resonance crossing experiments}
\subsection{Non-Markovian nonlinear Schr\"odinger equation}
Inserting the two-channel Hamiltonian (\ref{H_2B}) into 
Eq.~(\ref{pairfunction}) implies that the pair function has a closed channel 
component $\Phi_\mathrm{cl}(\mathbf{x},\mathbf{y},t)$, proportional to
$\phi_\mathrm{res}(r)$, and a long range component 
$\Phi_\mathrm{bg}(\mathbf{x},\mathbf{y},t)$ in the open channel. Only the 
$(F=2,m_F=-2)$ hyperfine state of $^{85}$Rb and the $(F=1,m_F=1)$ hyperfine 
state of $^{23}$Na have a mean field that we will denote simply by 
$\Psi(\mathbf{x},t)$. The dynamic equation (\ref{pairfunction}) can be solved 
formally in terms of the mean field and inserted into Eq.~(\ref{meanfield}) 
\cite{Koehler02,KGB02}. In the first order microscopic quantum dynamics 
approach the entire dynamics of the gas is then described by a single nonlinear
Schr\"{o}dinger equation \cite{Koehler02,KGB02,GKGTJ03,footnote:NLS}
\begin{align}
  \nonumber
  i\hbar\frac{\partial}{\partial t}\Psi(\mathbf{x},t)=& 
  \left[-\frac{\hbar^2}{2m}\nabla^2+V_\mathrm{trap}(\mathbf{x})\right]
  \Psi(\mathbf{x},t)\\
  &-\Psi^*(\mathbf{x},t)
  \int_{t_0}^\infty d\tau \
  \Psi^2(\mathbf{x},\tau)\frac{\partial}{\partial \tau}h(t,\tau),
  \label{NMNLSE}
\end{align}
with  
\begin{equation}
  h(t,\tau)=(2\pi\hbar)^3\langle0,\mathrm{bg}|V_\mathrm{2B}(B(t))\,
  U_\mathrm{2B}(t,\tau)|0,\mathrm{bg}\rangle\theta(t-\tau).
\end{equation}
Here $U_\mathrm{2B}(t,\tau)$ is the time evolution operator associated with
the magnetic field dependent Hamiltonian (\ref{H_2B}), $\theta(t-\tau)$ is the 
step function, which yields unity for $t>\tau$ and vanishes elsewhere, and 
$|0,\mathrm{bg}\rangle$ is the zero momentum plane wave of the relative motion 
of two atoms in the open channel. 
We note that Eq.~(\ref{NMNLSE}) accounts for the entire two-body collision 
dynamics including the continuum of modes above the dissociation threshold of 
the background scattering potential in a non-perturbative manner.

\subsection{Contributions beyond mean field approaches}
\label{subsec:Contributionsbeyondmeanfield}
The solution of Eq.~(\ref{NMNLSE}) determines the pair function, via 
Eq.~(\ref{pairfunction}), through its components in the open and closed 
channels:
\begin{equation}
  |\Phi(t)\rangle=|\mathrm{bg}\rangle \Phi_\mathrm{bg}(t)+
  |\mathrm{cl}\rangle \Phi_\mathrm{cl}(t).
\end{equation}
A derivation \cite{KGB02,GKGTJ03} beyond the scope of this paper shows that the
pair function determines the components of the density matrix of the 
non-condensed fraction in the open and closed channel to be
\begin{align}
  \Gamma_\mathrm{bg}(\mathbf{x},\mathbf{x}',t)=
  \int d\mathbf{y} \ \Phi_\mathrm{bg}(\mathbf{x},\mathbf{y},t)
  \Phi_\mathrm{bg}^*(\mathbf{x}',\mathbf{y},t)
  \label{Gammabg}
\end{align}
and
\begin{align}
  \Gamma_\mathrm{cl}(\mathbf{x},\mathbf{x}',t)=
  \int d\mathbf{y} \ \Phi_\mathrm{cl}(\mathbf{x},\mathbf{y},t)
  \Phi_\mathrm{cl}^*(\mathbf{x}',\mathbf{y},t),
  \label{Gammacl}  
\end{align}
respectively. The density of the gas is thus given by 
\begin{equation}
  n(\mathbf{x},t)=\Gamma_\mathrm{bg}(\mathbf{x},\mathbf{x},t)
  +\Gamma_\mathrm{cl}(\mathbf{x},\mathbf{x},t)
  +|\Psi(\mathbf{x},t)|^2,
\end{equation} 
and the total number of atoms $N=\int d\mathbf{x} \ n(\mathbf{x},t)$ 
is a constant of motion \cite{KGB02,GKGTJ03}.

In general, the solutions of the coupled equations (\ref{meanfield}) and 
(\ref{pairfunction}) 
depend not only on the initial condensate wave function $\Psi(\mathbf{x},t_0)$ 
and but also on the initial pair function $\Phi(t_0)$, which has so far been 
neglected in Eq.~(\ref{NMNLSE}). These initial conditions are determined by 
the experimental situation. For Feshbach resonance crossing experiments 
with $^{23}$Na it appears reasonable to require the condensate density to be 
stationary in the absence of any variation of the magnetic field strength. 
To find the stationary solution of the coupled equations (\ref{meanfield}) 
and (\ref{pairfunction}), we assume that the magnetic field strength at time 
$t_0$ is sufficiently far away from the Feshbach resonance that the binary 
dynamics is determined entirely by the background scattering. 
The closed channel component of the initial pair function can then be 
neglected, and for its open channel component $\Phi_\mathrm{bg}$ as well as 
for the mean field $\Psi$ we choose the {\em ansatz}\/:
\begin{align}
  \label{meanfieldstat}
  \Psi(\mathbf{R},t)&=\sqrt{n_0(\mathbf{R})}e^{-i\mu_0(t-t_0)/\hbar},\\ 
\Phi_\mathrm{bg}(\mathbf{R},\mathbf{r},t)&=\Phi_\mathrm{bg}(\mathbf{R},\mathbf{r},t_0) 
e^{-i2\mu_0(t-t_0)/\hbar}.
\end{align}
Here $\mathbf{R}=(\mathbf{x}+\mathbf{y})/2$ and 
$\mathbf{r}=\mathbf{y}-\mathbf{x}$ are centre of mass and relative 
coordinates, respectively, $n_0$ is the equilibrium condensate density, and
$\mu_0$ can be interpreted as the chemical potential.
The formal solution of Eq.~(\ref{pairfunction}) then provides the general 
form of the stationary pair function 
\begin{align}
  \Phi_\mathrm{bg}(\mathbf{R},\mathbf{r},t_0)=n_0(\mathbf{R})
  (2\pi\hbar)^{3/2}\mathrm{Re}
  \left[
    \langle\mathbf{r}|G_\mathrm{bg}(2\mu_0+i0)V_\mathrm{bg}|0\rangle
    \right],
\end{align}
where
\begin{equation} 
  G_\mathrm{bg}(2\mu_0+i0)=
  \left[
    2\mu_0+i0+\hbar^2\nabla^2/m-V_\mathrm{bg}
    \right]^{-1}
\end{equation} 
denotes the energy dependent Green's function associated with the background 
scattering potential $V_\mathrm{bg}$. Inserting the stationary {\em ansatz} 
of Eq.~(\ref{meanfieldstat}) for the mean field into Eq.~(\ref{meanfield}) 
then yields the stationary nonlinear Schr\"odinger equation:
\begin{align}
  \nonumber
  \mu_0\sqrt{n_0(\mathbf{x})}=&\left[
  -\frac{\hbar^2}{2m}\nabla^2+V_\mathrm{trap}(\mathbf{x})\right]
  \sqrt{n_0(\mathbf{x})}\\
  &
  +[n_0(\mathbf{x})]^{3/2} (2\pi\hbar)^3 \mathrm{Re}
  \left[\langle 0|T_\mathrm{bg}(2\mu_0+i0)
  |0\rangle\right].
  \label{mu0}
\end{align}
Here $T_\mathrm{bg}(2\mu_0+i0)$ is the 
energy dependent $T$ matrix associated with the background scattering potential
$V_\mathrm{bg}$ (see, e.g., Ref.~\cite{Koehler02}). Evaluating the $T$ matrix
at the energy $2\mu_0\to 0$, Eq.~(\ref{mu0}) recovers the usual 
Gross-Pitaevskii equation in the dilute gas limit 
($n_\mathrm{pk} a_\mathrm{bg}^3\ll 1$, where $n_\mathrm{pk}$ is the peak 
density), which then determines the chemical potential $\mu_0$ as well as 
the stationary condensate density $n_0(\mathbf{x})$. In the experiments
with $^{23}$Na \cite{Stenger99} the peak condensate density is on the order
of $n_\mathrm{pk}\approx 1.5\times 10^{15}\,$cm$^{-3}$ and including the 
initial pair function influences the dynamics on the time scales of the 
passage across the Feshbach resonances. We have thus included the initial 
pair function in Eq.~(\ref{NMNLSE}) as explained in Ref.~\cite{Koehler02}. 
In the experiments with $^{85}$Rb 
the peak density $n_\mathrm{pk}\approx 2.2\times 10^{12} \, \mathrm{cm}^{-3}$ 
turns out to be sufficiently low for initial pair correlations to be 
negligible. 

\subsection{Two level mean field approach}
The two level mean field approach of 
Refs.~\cite{Drummond98,Timmermans99,Abeelen99}
can be derived from the first order microscopic quantum dynamics approach
in Eqs.~(\ref{meanfield}) and (\ref{pairfunction}) with the {\em ansatz}
\begin{equation}
  \Phi_\mathrm{cl}(\mathbf{R},\mathbf{r},t)=\sqrt{2}\phi_\mathrm{res}(r)
  \Psi_\mathrm{res}(\mathbf{R},t)
  \label{poleapproxPhicl}
\end{equation}
for the closed channel component of the pair function in centre of mass and
relative coordinates. Here $\phi_\mathrm{res}(r)$ is the closed channel 
resonance level. A derivation \cite{GKGTJ03} beyond the scope of this
paper then shows that Eqs.~(\ref{meanfield}) and (\ref{pairfunction}) yield
the coupled equations
\begin{align}
  \nonumber
  i\hbar\frac{\partial}{\partial t}\Psi(\mathbf{x},t)=&
  \left[
    -\frac{\hbar^2}{2m}\nabla^2+\frac{m}{2}\omega_\mathrm{ho}^2|\mathbf{x}|^2
    \right]\Psi(\mathbf{x},t)\\
  \nonumber
  &+g_\mathrm{bg}|\Psi(\mathbf{x},t)|^2\Psi(\mathbf{x},t)\\
  &+g_\mathrm{res}\Psi^*(\mathbf{x},t)\sqrt{2}
  \Psi_\mathrm{res}(\mathbf{x},t)
  \label{2levelMFPsi}
\end{align}
for the condensate mean field, and
\begin{align}
  \nonumber
  i\hbar\frac{\partial}{\partial t}\Psi_\mathrm{res}(\mathbf{R},t)=&
  \left[
    -\frac{\hbar^2}{4m}\nabla^2+\frac{2m}{2}\omega_\mathrm{ho}^2|\mathbf{R}|^2
    \right]
  \Psi_\mathrm{res}(\mathbf{R},t)\\
  \nonumber
  &
  +\left[\frac{d E_\mathrm{res}}{d B}(B_\mathrm{res})\right]
  [B(t)-B_0]
  \Psi_\mathrm{res}(\mathbf{R},t)\\ 
  &+\frac{1}{\sqrt{2}}g_\mathrm{res}\Psi^2(\mathbf{R},t)
  \label{2levelMFPsires}
\end{align}
for the mean field associated with the resonance level in the Markov 
approximation. Here we have assumed that the atom trap is spherical and that
the closed channel resonance state experiences twice the trap potential of the 
atoms. The coupling constants $g_\mathrm{bg}$ and $g_\mathrm{res}$ are given by
$g_\mathrm{bg}=4\pi\hbar^2 a_\mathrm{bg}/m$ and 
$g_\mathrm{res}=(2\pi\hbar)^{3/2}
\langle\phi_\mathrm{res}|W|\phi_0^{(+)}\rangle$, where $a_\mathrm{bg}$ is the 
background scattering length, and $\phi_0^{(+)}(\mathbf{r})$ is the zero energy
wave function associated with the background scattering \cite{GKGTJ03}. The
physical picture underlying the two level mean field approach 
(see, e.g., Ref.~\cite{JTK03}) can be described
in terms of a linear curve crossing between the energy $E_\mathrm{res}(B)$
associated with the closed channel resonance level $\phi_\mathrm{res}(r)$
and the zero energy level of two condensate atoms (see
Fig.~\ref{fig:E_b85Rb}). This picture, however, neglects all the other
modes of the two-body background scattering
continuum and, consequently, does not recover the nature of the
highest excited vibrational diatomic bound state
\cite{TKTGPSJKB03}. Even more importantly, in the context of this paper, the absence of continuum
modes also implies that the two level mean field approach is unsuited to
describe the energy distribution of the atoms lost from the condensate
in an upward ramp of the Feshbach resonance level.

\section{Comparison between different many-body approaches}
\label{sec:Comparisonmanybodyapproaches}
\subsection{Loss of condensate atoms in upward ramps of the resonance level}
Since, in accordance with Eqs.~(\ref{Gammabg}) and (\ref{Gammacl}), the 
non-condensate density stems directly from the pair function, the build-up of 
pair correlations is the dominant source of atom loss from the condensate. In 
general, a time dependent magnetic field can transfer pairs of trapped 
condensate atoms either into highly excited diatomic molecules or into the 
complementary part of the two-body energy spectrum that contains a quasi 
continuum of excited levels \cite{KGB02}. In the experiments in 
Refs.~\cite{Stenger99,Cornish00} molecular formation can be ruled out because 
the linear ramps start on the side of the Feshbach resonance where the 
scattering length is positive and cross the resonance in such a way that the 
highest excited multichannel vibrational bound state $\phi_\mathrm{b}$ 
ceases to exist during the
passage (see Fig.~\ref{fig:E_b85Rb}) \cite{footnote:phi_res}. The remaining 
source of atom loss from the condensate is the formation of correlated pairs 
of atoms in initially unoccupied excited energy levels. As the time dependent 
homogeneous magnetic field provides only energy but no momentum the centre of 
mass motion of these pairs is limited by the initial momentum spread of the 
Bose-Einstein condensed gas. Their comparatively fast relative motion, however,
has been observed as a very dilute burst of atoms, e.g., in 
Ref.~\cite{Donley02}, and the typical scale of experimental single particle 
kinetic energies of $E_\mathrm{kin}/k_\mathrm{B}\sim 150 \ \mathrm{nK}$ has 
been confirmed with the first order microscopic quantum dynamics approach of 
this work \cite{KGB02}. We note that the two level mean field approach can 
only describe the transfer of condensate atoms into the resonance level
$\phi_\mathrm{res}(r)$. The subsequent decay of this metastable state has 
been modelled in Ref.~\cite{Abeelen99} in terms of a constant decay width. 
Corrections to this approach, which account for the energy dependence of this 
decay width have been provided in Refs.~\cite{Mukaiyama03,GKGTJ03}.

\subsection{Experimental conditions}
We have solved Eq.~(\ref{NMNLSE}) and Eqs.~(\ref{2levelMFPsi}) and 
(\ref{2levelMFPsires}) using a cylindrical trap potential with the radial 
(axial) trap frequencies of $\nu_{r}=150 \, \mathrm{Hz}$ 
($\nu_z=1500 \, \mathrm{Hz}$) for the sodium experiments \cite{Stenger99}, and 
of $\nu_{r}=17.5 \, \mathrm{Hz}$ ($\nu_z=6.8 \, \mathrm{Hz}$) in the case of 
$^{85}$Rb \cite{Cornish00}. As the initial condensate mean field we have used 
the ground state of the Gross-Pitaevskii equation with $N=900000$ $^{23}$Na 
atoms \cite{Stenger99} and $N=1594$ $^{85}$Rb atoms \cite{Cornish}, 
respectively. The initial scattering length was chosen to be the background 
scattering length $a_\mathrm{bg}$ for the $^{23}$Na atoms 
(cf.~Table \ref{tab:resparams}), and $a=228 \, a_\mathrm{Bohr}$ in the case
of $^{85}$Rb, corresponding to the experimental evaporation field strength of 
about 162.2 G. Under these initial conditions the condensate is weakly 
interacting ($^{23}$Na: $n_\mathrm{pk}a^3\approx 6\times 10^{-5}$ with a peak
density of $n_\mathrm{pk}\approx 1.5\times 10^{15} \, \mathrm{cm}^{-3}$; 
$^{85}$Rb: $n_\mathrm{pk}a^3\approx 10^{-6}$ with a peak density of
$n_\mathrm{pk}\approx 2.2\times 10^{12} \, \mathrm{cm}^{-3}$). 

Starting from 162.2 G, in the case of $^{85}$Rb, we have studied linear 
ramps of the magnetic field strength across the 155 G Feshbach resonance 
down to 132 G \cite{Cornish00}. In the course of these studies we have 
varied the inverse ramp speeds within the experimental range \cite{Cornish00} 
between $1.2 \, \mu\mathrm{s/G}$ and $200 \, \mu\mathrm{s/G}$.
The sodium ramps where extended over much larger timescales in the experiments 
\cite{Stenger99,Ketterleprivcomm}. We have therefore chosen the ramp intervals 
in such a way that for each ramp speed the detuning $|E_\mathrm{res}(B)|$ of 
the closed channel resonance state was sufficiently large at the initial and 
final times that increasing the ramp intervals did not significantly change the
total loss of atoms from the condensate.

\subsection{Predictions of the different many-body approaches}
\begin{figure}[tb]
\begin{center}
\includegraphics[width=\columnwidth,clip]{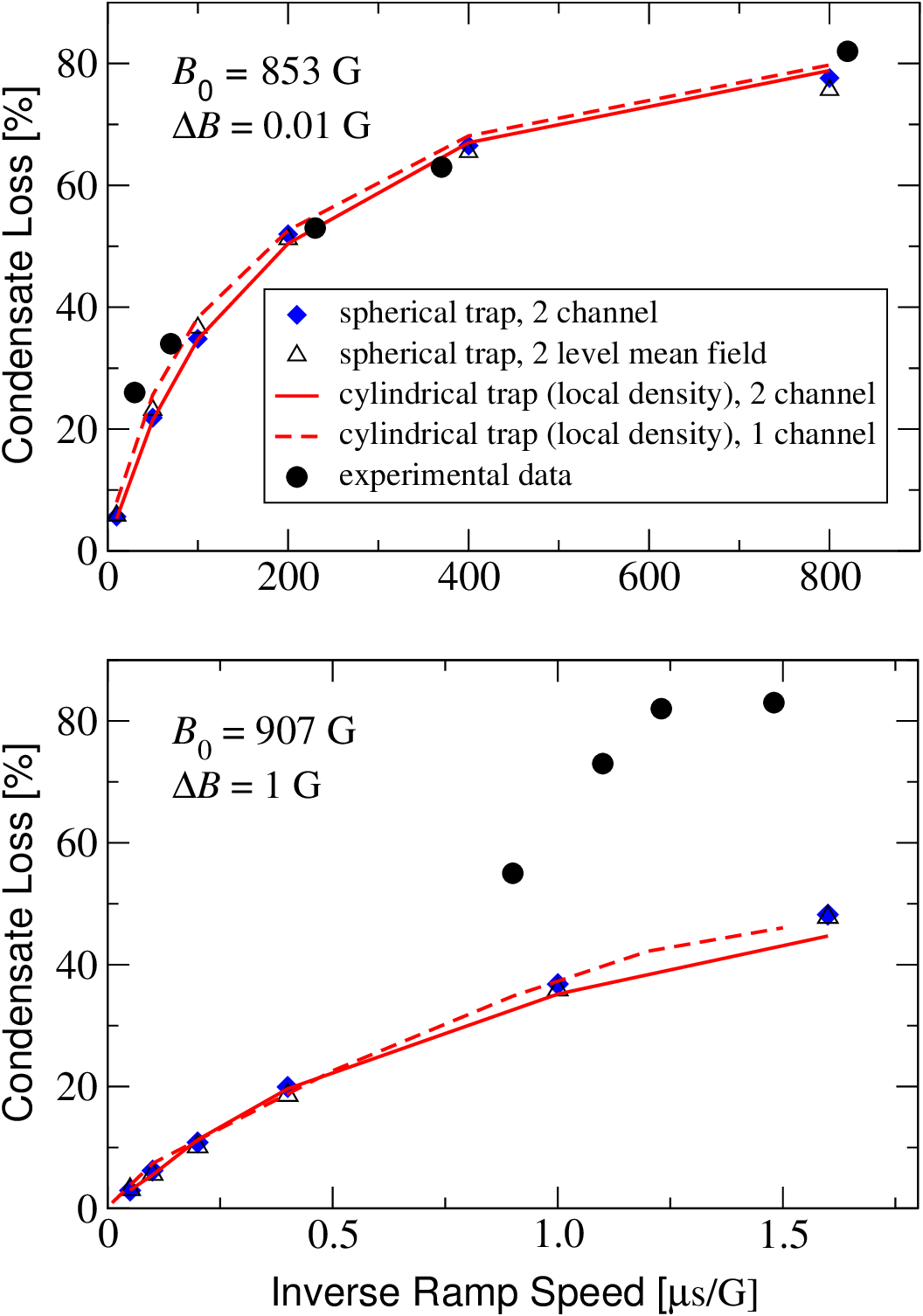}
\caption{
  \label{lossNa}
Loss of atoms from a $^{23}$Na condensate at the end of a linear magnetic 
field ramp across the 853 G and 907 G Feshbach resonances, 
respectively, as a function of the inverse ramp speed. The filled circles are 
the experimental data from Ref.~\protect\cite{Stenger99}. The diamonds are 
determined from Eq.~(\ref{NMNLSE}) using a spherical trap potential with the 
mean experimental trap frequency. The solid curve represents the calculation
in the local density approximation, solving Eq.~(\ref{NMNLSE}) for uniform 
gases, with densities weighted according to the initial distribution in a 
cylindrical trap. The dashed curve shows the corresponding result from a 
single channel calculation as described in Refs.~\protect\cite{KGB02,GKGTJ03}. 
The triangles correspond to the predictions of the two level mean field 
approach of Eqs.~(\ref{2levelMFPsi}) and (\ref{2levelMFPsires}) with a 
spherical trap potential.}
\end{center}
\end{figure}

Figure \ref{lossNa} shows the theoretical condensate loss curves 
$(N-\int d\mathbf{x} |\Psi(\mathbf{x},t)|^2)/N$ in dependence on the 
inverse ramp speed for the $^{23}$Na Feshbach resonances \cite{Na907G}, 
together with the experimental data from Ref.~\cite{Stenger99}. In the course 
of these studies, we have solved Eq.~(\ref{NMNLSE}) with a variety of different
methods. These methods involve complete solutions of Eq.~(\ref{NMNLSE}) 
including the potential of a spherical trap with the mean frequency 
$\nu_\mathrm{ho}=\sqrt[3]{\nu_{r}^2\nu_z}=697\,$Hz, as well as the local 
density approximation, which consists in solving Eq.~(\ref{NMNLSE}) in the 
homogeneous gas limit with densities weighted according to the initial density 
distribution associated with the realistic cylindrical atom trap. We have 
performed these calculations including the two-body time evolution associated 
with the full two channel Hamiltonian (\ref{H_2B}) and with the single channel 
approach described in Refs.~\cite{KGB02,GKGTJ03}. The single channel approach 
includes the time evolution of two atoms in the open channel with the 
background scattering potential adjusted in such a way that it recovers, at 
each magnetic field strength, the scattering length in Eq.~(\ref{scattlength}) 
as well as the Gribakin and Flambaum estimate of the binding energy $E_b$ in 
Eq.~(\ref{E_bGF}). We note that the single channel approach does not 
explicitly include the closed channel resonance state. For comparison, we have 
also solved Eqs.~(\ref{2levelMFPsi}) and (\ref{2levelMFPsires}) of the two 
level mean field approach using a spherical trap potential with the mean 
frequency of the cylindrical trap in the experiment \cite{Stenger99}.

The comparison between the theoretical predictions shows that although the 
description of the underlying microscopic collision physics differs decisively 
among the approaches, the predicted total loss of condensate atoms is always 
virtually the same. The remnant condensate fractions are therefore largely 
insensitive not only to phenomena related to the inhomogeneity of realistic 
Bose-Einstein condensates, like collective excitations, but also to the two 
channel nature of the two-body energy states at magnetic field strengths in 
the vicinity of the Feshbach resonances. In particular, the applicability of 
the single channel approach of Refs.~\cite{KGB02,GKGTJ03} appears surprising  
because the underlying single channel Gribakin and Flambaum estimate 
\cite{Gribakin93} of the near resonant binding energy $E_\mathrm{b}$ in 
Eq.~(\ref{E_bGF}) is rather poor in the case of the $^{23}$Na Feshbach 
resonances (cf.~Fig.~\ref{fig:E_bNa907G}). The comparison between theory and 
experiment suggests good agreement for the 853 G Feshbach resonance of 
$^{23}$Na but there are significant deviations for the broader 907 G 
resonance. Similar observations were reported also in Ref.~\cite{Abeelen99}.

\begin{figure}[tb]
\begin{center}
\includegraphics[width=\columnwidth,clip]{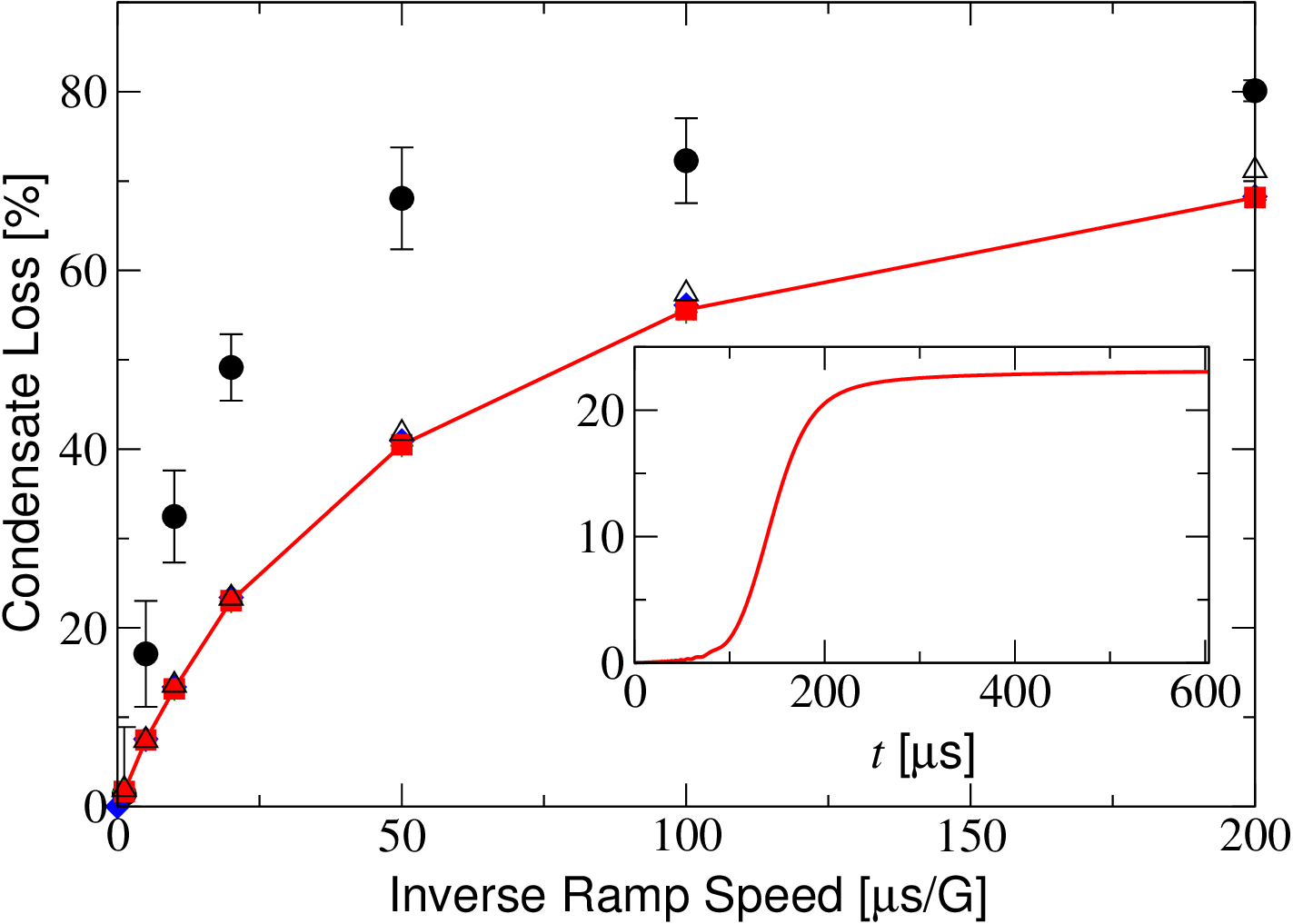}
\caption{\label{lossRb}
Loss of atoms from a $^{85}$Rb condensate at the end of a linear ramp of the
magnetic field strength across the 155 G Feshbach resonance as a function 
of the inverse ramp speed. The circles with error bars are the experimental 
data from Ref.~\protect\cite{Cornish00}. The squares are determined from 
Eq.~(\ref{NMNLSE}) using a cylindrical trap potential with the experimental 
trap frequencies. The diamonds are obtained with a spherical trap potential 
that accounts only for the mean trap frequency and the stars represent a local 
density approximation for the cylindrical trap. The triangles correspond to 
the two level mean field approach of Eqs.~(\ref{2levelMFPsi}) and 
(\ref{2levelMFPsires}) in a spherical trap. The inset shows the dynamics of 
the loss of condensate atoms in the cylindrical trap for the 
$20 \, \mu$s/G ramp.}
\end{center}
\end{figure}

Figure \ref{lossRb} shows the analogous theoretical predictions of the 
different approaches with respect to the loss of condensate atoms 
$(N-\int d\mathbf{x} \ |\Psi(\mathbf{x},t)|^2)/N$ in a gas of $^{85}$Rb, 
together with the associated experimental data from Ref.~\cite{Cornish00}, in
dependence on the inverse ramp speed. Also in this case, all many-body 
approaches, including the two level mean field model, yield virtually the same 
results. This is particularly remarkable because the physical picture
of a single linear curve crossing \cite{JTK03}, underlying the two level mean field 
approach, is not applicable to the experiments \cite{Cornish00} 
(cf.~Subsection \ref{subsec:differencesbetweenresonances}). The predicted 
monotonic curves in Fig.~\ref{lossRb} follow the experimental trend but 
they consistently underestimate the observed losses in Ref.~\cite{Cornish00}.

\section{Universality of molecular dissociation spectra}
\label{sec:dissociationspectra}
Upward ramps of Feshbach resonance levels also play an important role in the
detection of ultracold molecules produced in atomic Bose-Einstein condensates,
as they are frequently used for the purpose of molecular dissociation upon 
passage across the resonance \cite{Herbig03,Duerr03,Mukaiyama03,Duerr04}. The 
dissociation mechanism can be readily understood from the binding energies in
Fig.~\ref{fig:E_b85Rb} for the example of $^{85}$Rb; tuning the magnetic field 
strength across the resonance position from the high field side adiabatically 
transfers the bound state $\phi_\mathrm{b}$ into the zero energy state of two
asymptotically free atoms in the open channel. Starting from the molecular 
bound state $\phi_\mathrm{b}(B_i)$ at the magnetic field strength $B_i$, 
the probability of detecting a free pair of atoms with a relative energy 
between $E$ and $E+dE$ is given by \cite{GKGTJ03}:
\begin{equation}
  n(E)dE=p^2dp\int d\Omega_\mathbf{p} \ \left|\langle\phi_\mathbf{p}(B_f)|
  U_\mathrm{2B}(t_f,t_i)|\phi_\mathrm{b}(B_i)\rangle\right|^2.
  \label{dissociationspectrumgeneral}
\end{equation}
Here $\phi_\mathbf{p}(B_f)$ is the continuum energy state of the two-body two 
channel Hamiltonian (\ref{H_2B}) \cite{GKGTJ03}, at the final magnetic field 
strength $B_f$ of the ramp, which is associated with the final relative 
momentum $\mathbf{p}$ of the atom pair, and $d\Omega_\mathbf{p}$ denotes the 
angular component of $d\mathbf{p}$. We note that the relative kinetic energy 
of the atoms is related to their relative momentum by $E=p^2/m$.

\begin{figure}[tb]
\begin{center}
\includegraphics[width=\columnwidth,clip]{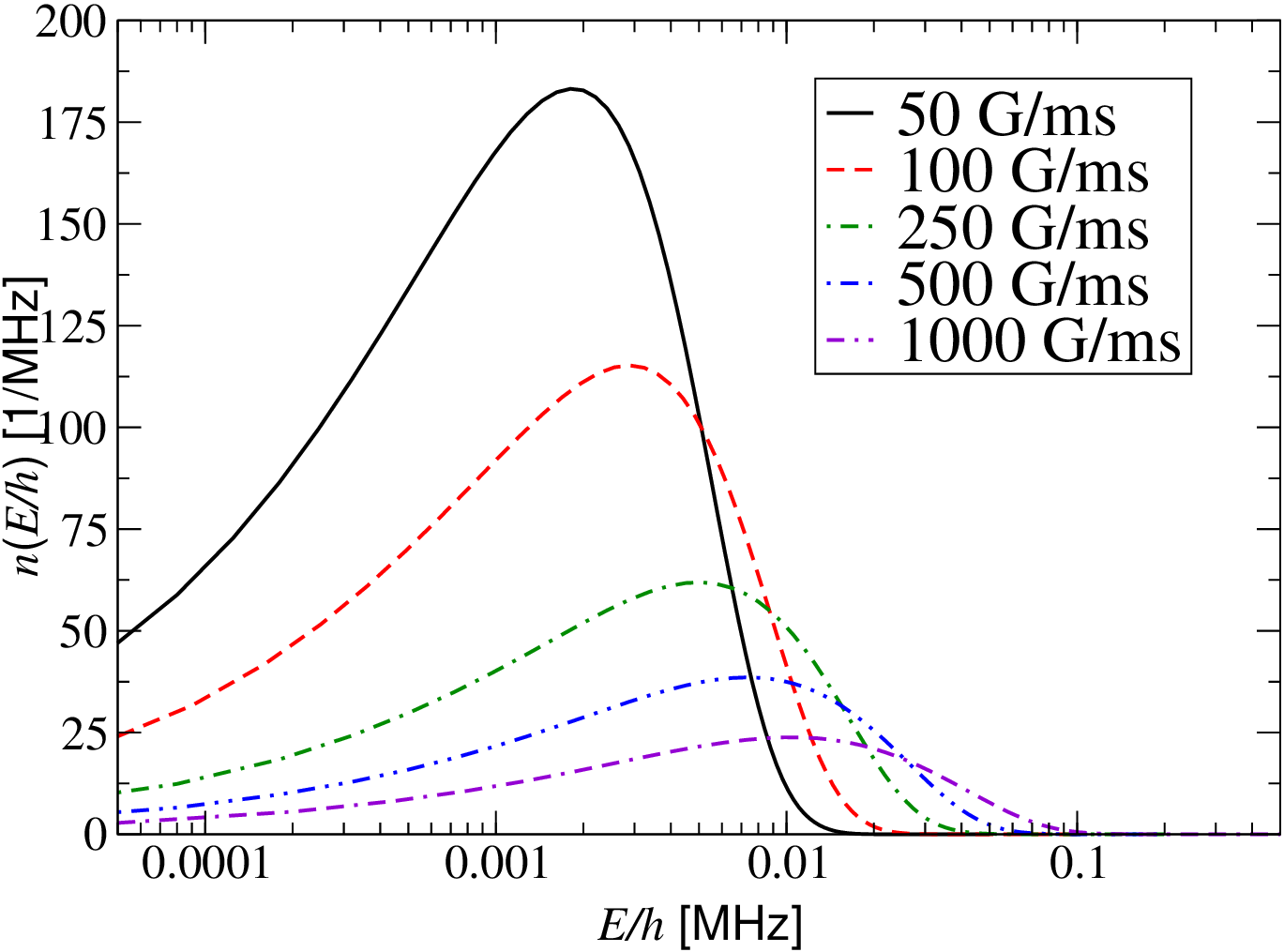}
\caption{Dissociation spectra of the highest excited vibrational bound state
    of $^{85}$Rb$_2$ as a function of the relative energy of the fragments. 
    The speeds of the linear ramps across the 155 G Feshbach resonance 
    were varied between 50 G/ms (solid curve) and 1000 G/ms 
    (double dashed dotted curve). The ramps started from the experimental 
    evaporation magnetic field strength of $B_i=162.2$ G \cite{Cornish00}
    down to a final magnetic field strength $B_f<B_0=154.9$ G, which was 
    chosen sufficiently small that at each ramp speed the spectrum was 
    insensitive to variations of $B_f$. The exact calculations based on  
    Eq.~(\ref{dissociationspectrumgeneral}) were performed with the two 
    channel Hamiltonian (\ref{H_2B}). We note that the relative kinetic 
    energies of the molecular fragments are given on a logarithmic scale.
  \label{fig:85Rbspectra}}
\end{center}
\end{figure}

Figure \ref{fig:85Rbspectra} shows the dissociation spectra $n(E)$ as obtained
from Eq.~(\ref{dissociationspectrumgeneral}) for the dissociation of
$^{85}$Rb$_2$ Feshbach molecules in the highest excited multichannel 
vibrational state $\phi_\mathrm{b}$ upon passage across the 155 G Feshbach
resonance. In these calculations the initial magnetic field strength was chosen
as the experimental evaporation magnetic field strength of $B_i=162.2$ G
reported in Ref.~\cite{Cornish00}. We have varied the speeds of the linear
ramps from 50 G/ms to 1000 G/ms, which illustrates that the 
dissociation spectra become broader in the energies with increasing ramp 
speeds.

The mean kinetic energies 
\begin{equation} 
  \langle E_\mathrm{kin}\rangle 
  =\frac{1}{2}\int_0^\infty dE \ E n(E)
  \label{definitionmeankineticenergies}
\end{equation}
of the molecular fragments after the dissociation characterise the speed of 
expansion of the gas of molecular fragments before the detection in related 
experiments \cite{Herbig03,Duerr03,Mukaiyama03,Duerr04}. We note that the kinetic 
energy of a single atom is $E_\mathrm{kin}=p^2/2m$, which is half the energy 
of the relative motion of a pair. A Fermi Golden Rule argument 
\cite{Mukaiyama03} shows that $\langle E_\mathrm{kin}\rangle$ can be expressed 
in terms of physical parameters of a Feshbach resonance, of the ramp speed 
$|dB/dt|$ and of Euler's $\Gamma$ function \cite{Mukaiyama03,GKGTJ03}:
\begin{equation} 
  \langle E_\mathrm{kin}\rangle =\frac{1}{3}
  \left[
    \frac{3}{4}\sqrt{\frac{\hbar^2}{m}}
    \frac{\hbar|dB/dt|}{|a_\mathrm{bg}||\Delta B|}
    \right]^{2/3}\Gamma(2/3).
  \label{asymptoticmeankineticenergies}
\end{equation}
The derivation of Eq.~(\ref{asymptoticmeankineticenergies}) given in 
Ref.~\cite{Mukaiyama03} relies upon the linear curve crossing picture described
in Refs.~\cite{Mies00,JTK03}, which identifies the stable multichannel 
molecular bound state $\phi_\mathrm{b}$ with the closed channel resonance 
state $\phi_\mathrm{res}(r)$ below the dissociation threshold of the background
scattering potential. Above the threshold $\phi_\mathrm{res}(r)$ acquires
a decay width dependent on the energy $E_\mathrm{res}(B)$, which is
associated with $\phi_\mathrm{res}(r)$. This Fermi Golden Rule decay width 
leads to the dissociation of the molecules. 

\begin{figure}[tb]
\begin{center}
\includegraphics[width=\columnwidth,clip]{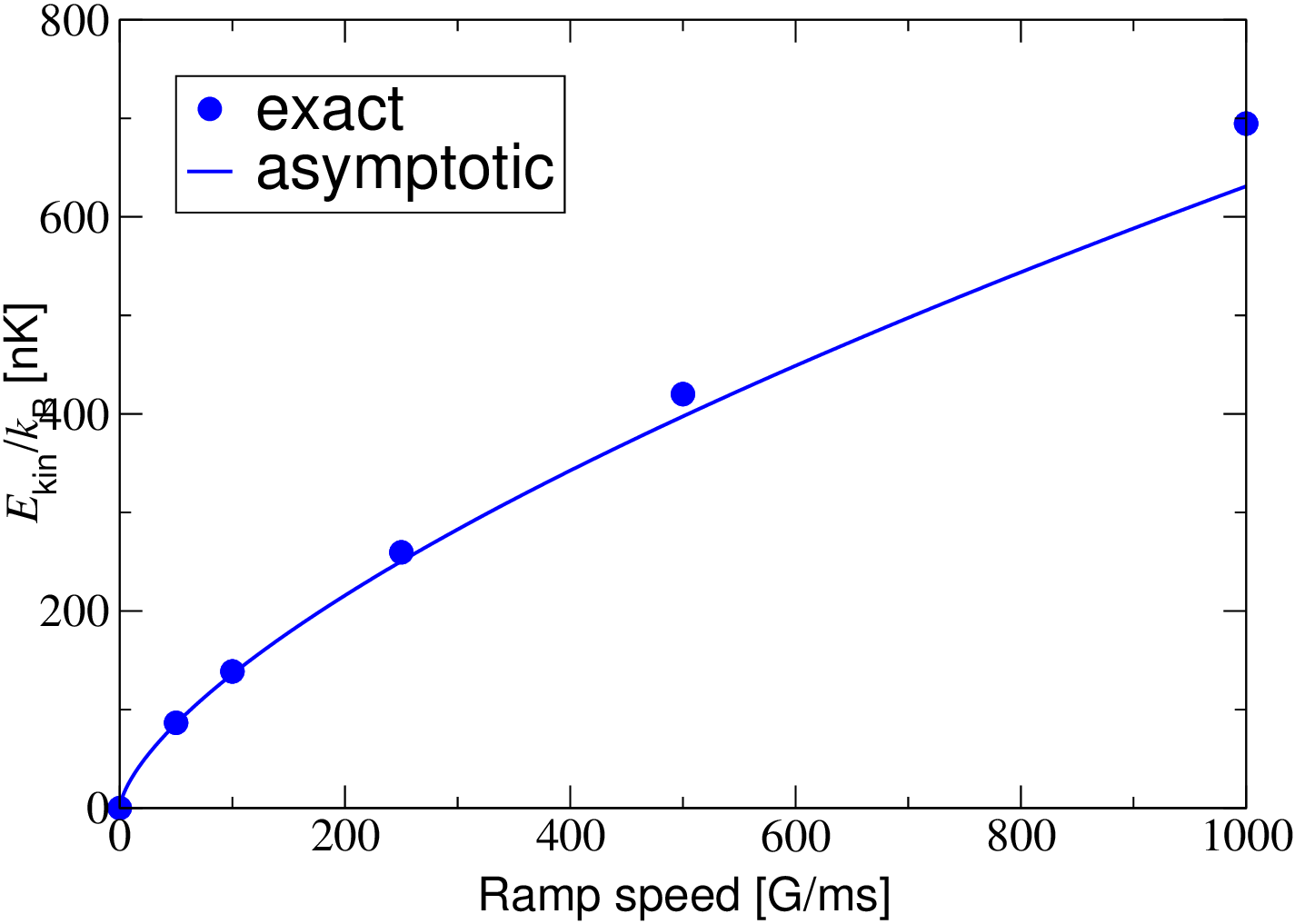}
\caption{Mean single particle kinetic energies of the molecular fragments 
    associated with the exact dissociation spectra obtained from 
    Eq.~(\ref{dissociationspectrumgeneral})
    in Fig.~\ref{fig:85Rbspectra} (circles) in dependence on the ramp 
    speed. The solid curve has been obtained from the asymptotic prediction
    in Eq.~(\ref{asymptoticmeankineticenergies}) for comparison. 
    The single particle kinetic energy is half the energy of the relative 
    motion of the fragments in Fig.~\ref{fig:85Rbspectra}.
  \label{fig:85Rbmeanenergies}}
\end{center}
\end{figure}

We expect Eq.~(\ref{asymptoticmeankineticenergies}) to be accurate provided 
that the initial and final magnetic field strengths of the linear ramp are 
sufficiently far from the resonance position. This has been demonstrated 
in the experiments of Ref.~\cite{Mukaiyama03}, in which $^{23}$Na$_2$ 
molecules in the highly excited multichannel vibrational bound state 
$\phi_\mathrm{b}$ were dissociated upon passage of the magnetic field strength 
across the 907 G Feshbach resonance. The derivation in 
Ref.~\cite{Mukaiyama03} is obviously inapplicable to $^{85}$Rb ramps in the 
experimental range of magnetic field strengths from 162.2 down to 132 G of 
Ref.~\cite{Cornish00} because the resonance level stays always above the 
threshold and the molecular bound state $\phi_\mathrm{b}$ has only a small 
admixture of $\phi_\mathrm{res}(r)$ ($<30$ \% in Fig~\ref{fig:populationcl}). 
Figure \ref{fig:85Rbmeanenergies} shows, however, that the analytic formula
(\ref{asymptoticmeankineticenergies}) is remarkably accurate. Small deviations 
between Eq.~(\ref{asymptoticmeankineticenergies}) and the exact mean energies
obtained from Eqs.~(\ref{dissociationspectrumgeneral}) and
(\ref{definitionmeankineticenergies}) only occur with increasing ramp speeds.
The molecular dissociation spectra thus exhibit the same insensitivity
to the exact microscopic binary collision physics as the remnant condensate
fractions in Section \ref{sec:Comparisonmanybodyapproaches}.

\section{Conclusions}
We have shown in Sections \ref{sec:Comparisonmanybodyapproaches} and
\ref{sec:dissociationspectra} that remnant condensate fractions or molecular
dissociation spectra associated with {\em linear} upward ramps of a Feshbach 
resonance level do not reveal the details of the underlying microscopic binary
collision physics. The formula (\ref{asymptoticmeankineticenergies}) for the
mean single particle kinetic energy of molecular fragments suggests 
that the Landau Zener parameter \cite{Mies00,GKGTJ03}
\begin{equation}
  \delta_\mathrm{LZ}=4\pi
  \frac{N-1}{\mathcal{V}}\frac{\hbar}{m}
  \frac{|a_\mathrm{bg}||\Delta B|}{|dB/dt|}
  \label{deltaLZ}
\end{equation}
determines the universal observables in Sections 
\ref{sec:Comparisonmanybodyapproaches} and \ref{sec:dissociationspectra}
when $|a_\mathrm{bg}|$, $|\Delta B|$ and $|dB/dt|$ are the parameters that can 
be varied independently. Here $N$ is the number of atoms and $\mathcal{V}$ is 
the volume of the gas. For $N=2$ atoms the Landau Zener coefficient exactly 
determines the final population in the zero energy level by
\begin{equation} 
  p_0=\exp(-2\pi\delta_\mathrm{LZ})
  \label{p0}
\end{equation}
in a two level linear curve crossing model of a Feshbach resonance
\cite{GKGTJ03,JTK03}, provided
that the linear ramp starts and ends asymptotically far from the crossing 
point $B_\mathrm{res}$ (see Fig.~\ref{fig:E_b85Rb}). The analysis of the two 
level mean field equations (\ref{2levelMFPsi}) and (\ref{2levelMFPsires}) 
of Ref.~\cite{GKGTJ03} reveals that the final populations of the remnant 
condensate and of the closed channel resonance state are indeed determined by 
$\delta_\mathrm{LZ}$, i.e.~the final populations are constant when the 
parameters $|a_\mathrm{bg}|$, $|\Delta B|$ and $|dB/dt|$ are varied in such a 
way that $\delta_\mathrm{LZ}$ is kept constant. In particular, the dependence 
of Eq.~(\ref{deltaLZ}) on the modulus of the ramp speed shows that the final 
populations are insensitive with respect to the ramp direction.

The analysis in Ref.~\cite{Demkov68} shows that Eq.~(\ref{p0}) determines the 
exact asymptotic population $p_0$ in the open channel ground state
even when the complete quasi continuum of modes in the open channel is taken 
into account in a linear curve crossing model of a Feshbach resonance. This
explains why including the full two channel time evolution operator in  
Eq.~(\ref{NMNLSE}) leads to the same results as the two level mean field 
approach for the $^{23}$Na Feshbach resonances in Section 
\ref{sec:Comparisonmanybodyapproaches}, given that phenomena related to the
non Markovian nature of Eq.~(\ref{NMNLSE}) are negligible. Our studies of the
155 G Feshbach resonance of $^{85}$Rb in Figs.~\ref{lossRb} and 
\ref{fig:85Rbmeanenergies} suggest that also the asymptotic criterion of 
applicability of the two-body Landau Zener approach can be violated to a large 
extent without any significant changes in the results. We note,
however, that the two level picture is not suited to describe the
interferometric experiments of
Refs.~\cite{Donley02,Claussen03}, which employ a sequence of magnetic
field pulses. In fact, the occupation of binary continuum modes,
observed in these experiments, is ruled out in any two level approach
from the outset and, in addition, the nature of the
highest excited vibrational diatomic bound state is not
recovered. This explains why the two level mean field approach fails
to predict the experimentally observed atom-molecule Ramsey fringes \cite{KGTKKB04}.

Although the Landau Zener coefficient in Eq.~(\ref{deltaLZ}) can be derived 
from a linear curve crossing model of a Feshbach resonance, it is independent 
of the slope $dE_\mathrm{res}/dB$ of the resonance and contains just the
parameters $a_\mathrm{bg}$ and $\Delta B$ of the Feshbach resonance, which
determine the scattering length in Eq.~(\ref{scattlength}). In fact, the inset 
of Fig.~\ref{lossRb} shows that the main loss of condensate atoms occurs in the
close vicinity of the Feshbach resonance, where all low energy scattering 
properties become universal \cite{TKTGPSJKB03,GKGTJ03}, i.e.~the associated 
physical observables just depend on the scattering length $a$. The universal 
parameters $a_\mathrm{bg}$ and $\Delta B$ as well as the inverse ramp speed 
that determine Eq.~(\ref{deltaLZ}) are also accounted for in the single channel
description of the first order microscopic quantum dynamics approach of 
Ref.~\cite{KGB02}. We have, indeed, confirmed that the single channel 
description also recovers the theoretical loss curves in Fig.~\ref{lossRb}, 
as it does in the case of the $^{23}$Na Feshbach resonances in 
Fig.~\ref{lossNa}. Consequently, our studies clearly reveal that the 
losses of condensate atoms in linear ramps of the magnetic field strength
across Feshbach resonances depend, to a large extent, just on the Landau Zener 
parameter. The two or multichannel nature of Feshbach resonances is 
virtually irrelevant to the loss curves in Figs.~\ref{lossNa} and \ref{lossRb}.

In summary, we have shown that physical observables, like the loss of 
condensate atoms or molecular dissociation spectra, associated with 
{\em linear} upward ramps of a Feshbach resonance level, are largely 
independent of the microscopic binary collision dynamics and depend just on 
the parameter $|a_\mathrm{bg}||\Delta B|/|dB/dt|$ and on the local densities 
of the atoms in the gas. Consequently, related experimental studies are 
inconclusive with respect to the details of the interatomic interactions.
The interferometric studies in Refs.~\cite{Donley02,Claussen03} and their
subsequent theoretical analysis in Refs.~\cite{Koehler02,Kokkelmans02,KGB02,KGTKKB04}
have demonstrated, however, that the universality of the physical quantities 
described in this paper is characteristic for {\em linear} ramps of the magnetic 
field strength. 
 
\acknowledgments 
We are particularly grateful to Simon Cornish for providing the experimental 
data of Ref.~\cite{Cornish00} to us. We thank Simon Gardiner, Paul Julienne, 
Keith Burnett, Simon Cornish, and Wolfgang Ketterle for many interesting 
discussions. This research has been supported through a E.C.~Marie Curie 
Fellowship under Contract no.~HPMF-CT-2002-02000 (K.G.), a University Research 
Fellowship of the Royal Society (T.K.), and by the Deutsche 
Forschungsgemeinschaft (T.G.).


\begin{thebibliography}{99}
\bibitem{footnote:upward}
We note that an upward sweep of the Feshbach resonance level in energy can imply a
downward sweep of the magnetic field strength. This is the case when
the difference in magnetic moment between the Feshbach resonance level
and a pair of atoms in the open channel is negative. This situation
occurs, e.g., for the 155 G Feshbach resonance of $^{85}$Rb (see Fig.~\ref{fig:E_b85Rb}).
\bibitem{footnote:inelastic}
Within the range of ramp speeds of Ref.~\cite{Cornish00} in which the 
remnant condensate was nearly completely depleted analogous measurements 
on a thermal $^{85}$Rb gas ($T=430 \, \mathrm{nK}$), with a comparable mean 
density, did not exhibit any significant losses \cite{Cornish00}. 
Consequently, we expect inelastic collisions to be negligible because these 
losses should otherwise be visible also in the thermal gas.
\bibitem{footnote:NLS}
We note that the elimination of the pair function 
conveniently reduces the spatial 
degrees of freedom from six dimensions in Eqs.~(\ref{meanfield}) 
and (\ref{pairfunction}) to three dimensions in Eq.~(\ref{NMNLSE}).
\bibitem{footnote:phi_res}
We note that $|\mathrm{cl}\rangle\phi_\mathrm{res}(r)$ is not a  
bound state of the two-body Hamiltonian (\ref{H_2B}). 
This state would thus dissociate.
\bibitem{Na907G}
For the 853 G resonance of $^{23}$Na we have used the theoretical value of 
$\Delta B=0.01\,$G \protect\cite{Stenger99} for the resonance width, which 
is a factor of 4 larger than the experimental width reported in 
Ref.~\protect\cite{Stenger99}.
\end{thebibliography}
\end{document}